\documentclass[prl,twocolumn,showpacs,amsmath,amssymb,superscriptaddress]
{revtex4}

\usepackage[dvips]{color}
\usepackage{graphicx}

\begin{document}

\title{Quantum glass phases in the disordered Bose-Hubbard model}
  
\author{ Pinaki Sengupta}
\affiliation{T-CNLS and NHMFL, Los Alamos National Laboratory,
Los Alamos, NM 87545, USA}
\affiliation{Department of
    Physics \& Astronomy, University of Southern California, Los Angeles,
    CA 90089, USA}
\author{Stephan Haas}
\affiliation{Department of
    Physics \& Astronomy, University of Southern California, Los Angeles,
    CA 90089, USA}

\begin{abstract}
The phase diagram of the Bose-Hubbard model in the presence of 
off-diagonal disorder is determined  using Quantum Monte Carlo 
simulations. A sequence of quantum glass phases 
intervene at the interface between the Mott insulating and the Superfluid
phases of the clean system. In addition to the standard Bose glass phase, 
the coexistence of gapless and gapped regions close to the Mott insulating 
phase leads to a novel Mott glass regime which is incompressible yet gapless. 
Numerical evidence for the properties of these phases is given in terms of
global (compressibility, superfluid stiffness) and local (compressibility,
momentum distribution) observables.  
\end{abstract}


\maketitle

The competition between disorder, interactions and commensurability 
in quantum many-body systems is known to produce
novel quantum glassy phases, characterized by a gapless spectrum
and by the absence of a global order parameter\cite{Fisher-1989}.
$^4$He adsorbed
on porous media\cite{Reppy-1995}, granular superconductors\cite{Goldman-1998}, 
disordered magnets\cite{Oosawa-2002} are but a few manifestations of localization
effects due to random potentials in interacting bosons.
In recent years, ultracold atomic gases in
magneto-optical traps have opened a new frontier in the study 
of strongly correlated systems, as unprecedented control over 
experimental parameters in these systems makes
them ideally suited for studying many-body phenomena.
Disorder can be generated in optical lattices by 
exposure to speckle lasers\cite{Horak-2000,Clement-2005}, 
incommensurate lattice-forming lasers
\cite{Damski-2003,Sanpera-2004,Fallani-2006}, 
and by other means\cite{Folman-2002}. The interplay between
disorder and interactions in trapped Bose-Einstein condensates  
has recently been explored experimentally in $^{87}$Rb, both in the
continuum\cite{Lye-2005,Fort-2005,Clement-2005} and in an 
optical lattice\cite{Fallani-2006,Schulte-2005}.

\begin{figure}
\includegraphics[width=8.3cm]{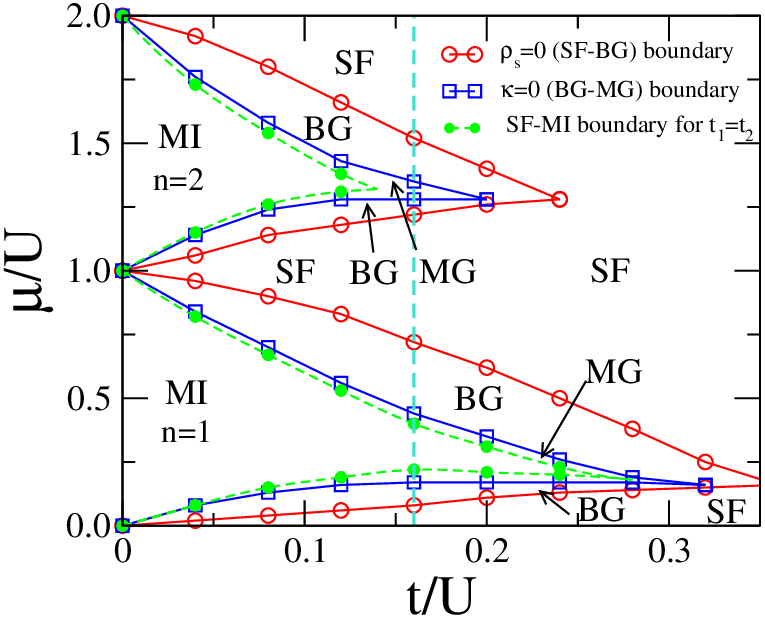}
\vskip1mm
\caption{(Color online) Phase diagram of the 1-D
BHM with spatially correlated off-diagonal disorder: 
$t_{i,i+1} = t/3$ and $t$ with equal probability. For comparison, 
the phase boundaries of the clean system ($t_{i,i+1}$ = $t$)
are shown with dashed lines. The 
phases are characterized by the stiffness $\rho_s$
and global compressibility $\kappa$: superfluid (SF) ($\rho_s>0,
\kappa>0$), Bose glass (BG) ($\rho_s=0, \kappa>0$), Mott insulator (MI)
($\rho_s=0, \kappa=0$). The Mott glass (MG) and MI phases have 
identical global properties and differ only in local 
properties.}
\label{fig:phases}
\end{figure}

Theoretically, the
effects of random potentials on interacting bosons in periodic lattices
have been studied using 
analytic\cite{Fisher-1989,deMartino-2005,Gavish-2005,Kuhn-2005}
and numerical techniques\cite{Scalettar-1991}. 
It is now well established that even infinitesimally small potential 
disorder can destroy the direct superfluid (SF) to Mott insulator (MI) 
transition in one dimension by introducing an intervening insulating,
but compressible, Bose glass (BG) phase\cite{Fisher-1989,Prokofev-1998}.
Surprisingly, while the effects
of potential disorder have been widely investigated, other kinds
of disorder, e.g. hopping or interaction strengths, have 
remained largely unexplored until recently, when it was demonstrated that 
these can lead to qualitatively quite different phenomena \cite{Wessel-2005,
Balabanyan-2005, Altman-2004}. In the 
quantum rotor model with off-diagonal disorder, 
the SF to MI transition takes place via an intermediate Mott glass (MG) 
phase \cite{Balabanyan-2005, Altman-2004}-- a unique incompressible, 
yet gapless, glassy regime that was first reported in disordered fermions 
with extended range interactions\cite{Giamarchi-2001}. It is thus of 
great interest 
to investigate whether such a phase also appears in the Bose-Hubbard model
(BHM), and if particle-hole symmetry is essential for the stabilization of 
such a MG.  

We use large-scale QMC simulations to study the effects of
off-diagonal disorder in the Bose-Hubbard model on a one-dimensional (1D)
lattice. We find that, in contrast to diagonal disorder,
the Mott lobes do not shrink in the presence of off-diagonal 
disorder. Instead, there is an extended BG phase separating the MI
lobes from the SF regime, and an additional MG regime emerges.

The Bose-Hubbard model is given by the Hamiltonian
\begin{equation}
H=\sum_{i=1}^L \left[ - t_{i,i+1}(b_{i+1}^{\dagger}b_i + h.c.)+ 
 {U\over 2}n_i(n_i - 1)- \mu n \right],
\label{eq:bhm}
\end{equation}  
where
$b_i^{\dagger}$ ( $b_i$) creates (annihilates) a boson 
at site $i$, $t_{i,i+1}$ is the hopping matrix element 
between sites $i$ and $i+1$, $U$ is the on-site interaction
strength, and $\mu$ is the chemical potential. Off-diagonal disorder is 
introduced in the form of a bimodal distribution of the hopping
matrix elements -- $t_{i,i+1}=t$ or $t/3$ with equal probability\cite{footnote1}.
The Stochastic Series Expansion (SSE)\cite{Anders-1999} method is used to 
simulate the model(\ref{eq:bhm}) in chains of length $L=64-256$. 
For each set of parameters, 200-1000 realizations of disorder are sampled.
To characterize the emerging phases, we compute the
superfluid stiffness and the global compressibility. In simulations 
employing updates that sample different winding number sectors, the 
stiffness is conveniently obtained from the fluctuations in the 
winding numbers of the world lines as  $\rho_s=[\langle W^2\rangle/2\beta]_{av}$,
where $[...]_{av}$ 
denotes averaging over multiple realizations of disorder.
The global compressibility, $\kappa$, is the energy cost of adding
a particle to the system. It is defined by $\kappa = 
\beta [\langle n^2\rangle - \langle n\rangle^2]_{av}$,
 where $n$ is the density of particles. 
 
At sufficiently large $U$,
local variations of the hopping amplitudes create a non-trivial landscape 
of higher-mobility domains (``lakes") that coexist with gapped, localized
regions. In order to tune the size of these lakes, we introduce spatially
correlated disorder. 
The Fourier filtering method\cite{corr-disorder} is 
used to generate a sequence of random
numbers, $\eta(i)$, with long-range algebraic correlation,
\begin{equation}
C(i-j) \equiv \langle \eta(i)\eta(j)\rangle \sim |i-j|^{-\alpha}.
\end{equation}
Without any loss of generality, we choose $\alpha$ = 0.3. 
This is then mapped onto a bimodal distribution of hopping integrals,
$t_{i,i+1}$, between sites $i$ and $i+1$ of a one-dimensional lattice. 
While the qualitative features are 
same for correlated and random disorder, spatially correlated disorder
favors extended  domains in the disorder realizations. This, 
in turn, allows for a more reliable 
identification of the phase boundaries of the unique
MG phase which differs from the MI only in terms of short-range properties. 
Its proper characterization thus
depends crucially on the sampling of finite-ranged domains. With 
uncorrelated disorder, typical domain sizes are smaller, which renders
the MG phase more difficult to detect numerically.

The results of the simulations are summarized in Fig.~\ref{fig:phases},
rendering a rich phase diagram.
In the absence of disorder, the ground state of the BHM 
is in either the SF or the MI phase.
The SF-MI transitions (dashed lines)
are marked by the simultaneous vanishing of the 
compressibility, $\kappa$, and stiffness, $\rho_s$, leading to the
well-known sequence of MI lobes, with integer fillings
\cite{mott,Scalettar-1990}, as $\mu/U$ is varied. In the presence 
of diagonal disorder, the two transitions -- $\rho_s=0$ and $\kappa=0$ -- 
decouple and a compressible ($\kappa > 0$), insulating 
($\rho_s=0$) BG phase appears. With off-diagonal disorder, 
an additional glassy phase -- the MG -- is realized. The
MG has global properties identical to the MI ($\rho_s=0, \kappa=0$ and
integer filling), but differs in local 
properties. In contrast to diagonal disorder,
the Mott lobes do not shrink with respect to the clean case.
There are no glassy phases in the atomic limit ($t/U \rightarrow 0$).
At finite $t/U$, the direct SF-MI transitions of
the clean limit are replaced by SF-BG-MG-MI sequences. 
Our results are consistent with all the transitions being continuous.

\begin{figure}
\includegraphics[width=8.3cm]{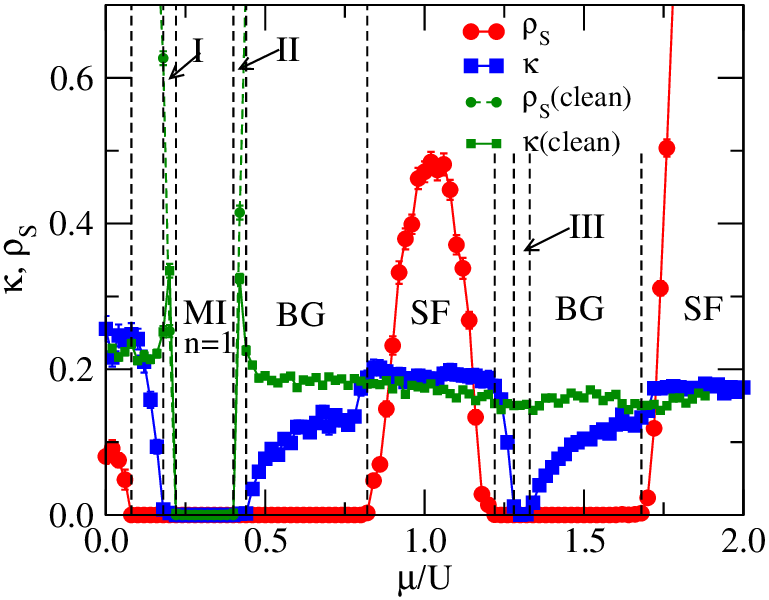}
\vskip1mm
\caption{(Color online) Stiffness ($\rho_s$) and 
compressibility ($\kappa$) as 
functions of the chemical potential ($\mu$) at constant $t/U$ =0.16 (cut along
the dashed line in Fig.\ref{fig:phases}). For comparison,
data of the clean system are also shown. In the absence of disorder, stiffness
and compressibility vanish simultaneously at the SF-MI boundary. 
In the presence of
disorder, the $\rho_s=0$ and the $\kappa=0$ transitions decouple, resulting
in the glassy phases BG ($\rho_s=0, \kappa>0$) and MG ($\rho_s=0, \kappa=0$).
The MG regimes I, II and III
differ from the MI in terms of local quantities. Note 
that there is a finite MG phase with $n=2$ ($1.28 < \mu/U < 1.34$) while
there is no corresponding MI phase.} 
\label{fig:rhokappa}
\end{figure}

Details of the simulations are illustrated in Fig.~\ref{fig:rhokappa}.
The global stiffness and compressibility are shown as a function of 
the chemical potential $\mu/U$ at constant $U/t=6.25$, i.e. along the dashed line
in Fig.~\ref{fig:phases}. The data for 
the clean system is shown for comparison. At small values of $\mu/U$,
the ground state is SF with $\rho_s > 0, \kappa > 0$. With
increasing $\mu/U$, the system passes through a series of phases, as
indicated in the figure. The first SF-BG transition is marked by the vanishing
of $\rho_s$, while $\kappa$ remains finite across the transition. This
coincides with the SF-MI boundary in the clean system with $t_{i,i+1}=t/3$
for all links.  In the BG phase, the ground state is a mixture of SF
and MI domains of all sizes -- the domains with $t_{i,i+1}=t/3$ are 
local MIs, 
 while those with $t_{i,i+1}=t$ are locally SF. The MI phase of the disordered system 
coincides with that in the clean limit with $t_{i,i+1}=t$. Of
particular interest are the regions marked I, II and III. For these
ranges of $\mu/U$, the global compressibility, $\kappa$, and the stiffness,
$\rho_s$, vanish identically. The ground state has integer filling
($n=1$ for regions I and II and $n=2$ for region III). The global
properties are thus identical to those of the MI phase. 
As shown next, unlike the MI phase, the ground state 
has locally compressible regions, 
and these ranges of $\mu/U$ can accordingly be identified as MG phases.
Since these regimes lie outside the Mott lobes of 
the clean system with $t_{i,i+1}=t$ (in particular, there is no $n=2$ 
MI phase for this value of $U/t$), the ground state consists of a mixture 
of SF ($t_{i,i+1}=t$) and MI ($t_{i,i+1}=t/3$) domains as in the BG phase,
but differs from the BG by being globally incompressible. 

\begin{figure}
\includegraphics[width=8.3cm]{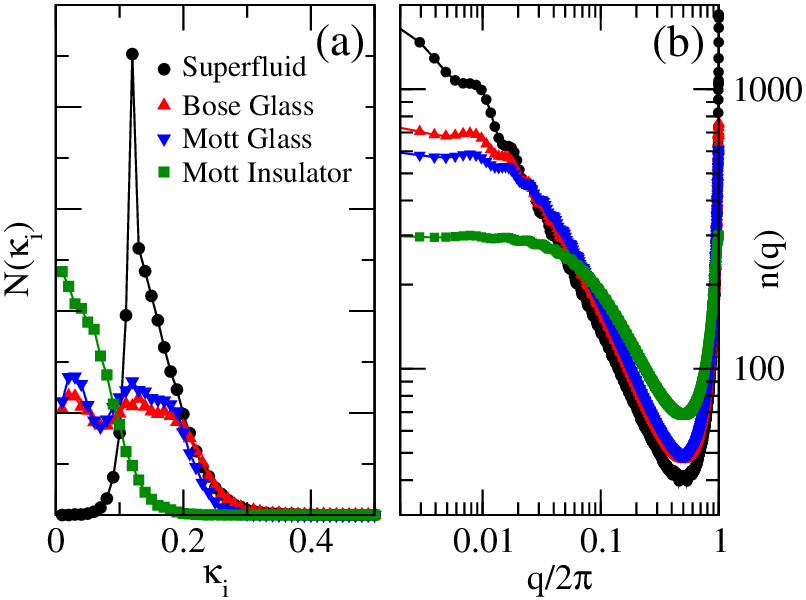}
\vskip1mm
\caption{(Color online) 
(a) The distribution of local (site) compressibilities in the different
phases averaged over 800 realizations of disorder. 
The double-peaked distribution
in the BG and MG phases indicate the presence of co-existing 
SF and MI 
domains.
(b) Momentum profiles of the corresponding ground states. The
BG and MG phases have similar momentum profiles, reflecting the similarity 
in their local structure.} 
\label{fig:localprops}
\end{figure}

Having explored the global properties, we focus 
on the local compressibility
and the momentum profile of the ground state in each phase. The local
compressibility at site $i$ is defined as the local number 
fluctuation, $\kappa_i = \beta [\langle n_i^2\rangle - \langle n_i\rangle^2]_{av}$,
 where $n_i$ is the particle density at site $i$.
The momentum distribution is obtained from the equal-time Green's
function,
\begin{equation}
n(q)={1\over N}\sum_{l,m}e^{-iq(r_l-r_m)}\langle b^{\dagger}_lb_m \rangle.
\end{equation}
$n(q \rightarrow 2\pi)$ measures the short-range coherence in the system.
The left panel of
Fig.~\ref{fig:localprops}
shows the distribution of the local compressibilities in
a lattice of length $N=128$, averaged over 800 disorder realizations.
In the MI phase, the distribution is peaked at small $\kappa_i$. Conversely,
in the SF phase it is peaked at a finite value. 
In the BG and MG phases, the distribution has a double-peaked 
structure, consistent with 
co-existing SF and MI domains in these phases\cite{footnote2}.
Fig.~\ref{fig:localprops}(b), shows momentum 
profiles in these phases. $n(q)$ is sharply peaked at $q=0$ in
the SF phase and exhibits only a weak broad maximum around $q=0$
in the MI phase. The behavior of the MG and BG phases is similar and
intermediate between the two limits -- there is a peak at $ q=0$ arising
from the SF domains, but the height is reduced due to the MI domains.
Thus the BG and MG phases have very similar short-range structure (both
phases consist of co-existing SF and MI domains), although their global 
responses are rather different, as documented by Fig.~\ref{fig:rhokappa}.
While the MG phase is globally incompressible, there exist a gapless
channel that allows the exchange of particles between adjacent SF and MI
domains, leading to its incompressible yet gapless character. Such
a mode is made possible due to the local variation of the 
kinetic energy of the bosons
in the different domains. This explains why the MG phase is not observed 
for purely diagonal disorder and is unique to off-diagonal disorder\cite{footnote3}.
The need to probe local properties also makes it difficult to detect this
phase in current optical lattice experiments, where, additionally,
domains of different phases co-exist. Current developments in local probes using 
selective microwave spectroscopy\cite{Campbell-2006} and the recently suggested 
method of using induced controlled interactions between a ``probe particle'' 
and the many-body state appear promising\cite{Giamarchi-2007}.

\begin{figure}
\includegraphics[width=8.3cm]{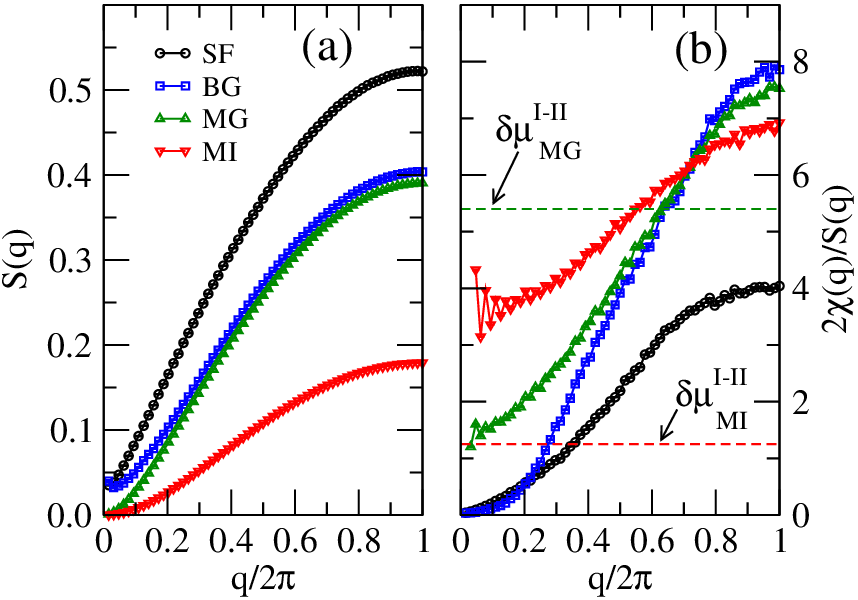}
\vskip1mm
\caption{(Color online) (a) Static structure factor, and (b) static
susceptibility for the diagonal correlation in the different
phases as a function of momentum. The effective potential difference
between type I ($t=1$) and type II ($t=1/3$) domains in the MG and 
MI phases are denoted by $\delta\mu^{I-II}_{MG}$ and $\delta\mu^{I-II}_{MI}$, respectively. 
Consistent with global compressibility measurements, $S(q)$ vanishes 
as $q\rightarrow 0$ in the MI and MG phases, but remains finite 
in the SF and BG phases. $\Delta_{MG} > 2\chi(q)/S(q)$ as $q\rightarrow 0$ 
in the MG phase, implying the existence of a gapless mode.} 
\label{fig:sq}
\end{figure}

To further characterize the various phases and probe 
the gapless nature of the MG phase, we study the static structure
factor, $S(q)$, for the density-density correlation 
\begin{equation}
S(q) = \left [\sum_{j,k} e^{-i\cdot q\cdot (r_j - r_k)}\langle n_jn_k\rangle\right ]_{av}
\end{equation}

and the associated susceptibility,
\begin{equation}
\chi(q)= \left [\int_0^{\beta}d\tau \sum_{j,k} e^{-i\cdot q\cdot (r_j - r_k)}\langle n_j(\tau)n_k(0)\rangle\right ]_{av}.
\end{equation}
A finite value of $S(q)$ as $q\rightarrow 0$
implies the phase is compressible, i.e., the 
energy gap towards adding or removing a particle vanishes in the
thermodynamic limit. Conversely, a vanishing $S(q)$ as $q\rightarrow 0$
is a signature of an incompressible phase with a gapped charge 
excitations. Additionally, the ratio
$2\chi(q)/S(q)$ as $q\rightarrow 0$ gives an upper bound for the 
charge excitation gap\cite{Hohenberg-1974}. 
Fig. \ref{fig:sq} shows the structure factor
and the susceptibility as a function of momentum for representative 
points in each of the different phases. Consistent with global 
compressibility measurements, $S(q)$ vanishes as $q\rightarrow 0$
in the MI and MG phases, whereas it remains finite for  SF and BG
ground states. The upper bound of the charge excitation gap,
$\Delta \equiv 2\chi(q)/S(q)$ as $q\rightarrow 0$, on the other hand, 
sheds light on the nature of the MG phase. Although
$\Delta_{MG}$ is finite, it is less than the effective chemical potential
difference $\delta\mu^{I-II}_{MG}$ between type I ($t=1$) and 
type II ($t=1/3$) domains. While there is 
a gap towards overall addition or removal of bosons, there exists
a gapless mode involving the transfer of particle across the boundaries
of the two types of domains which is driven by the effective potential energy
difference between adjacent sites at the domain boundaries. This is
in contrast to the MI phase where the effective potential difference 
$\delta\mu^{I-II}_{MI}$
between the two types of domains is less than $\Delta_{MI}$. This confirms
the existence of a gapless mode in the MG phase. 

In conclusion, using large-scale quantum Monte Carlo simulations, we have 
observed that off-diagonal disorder (random hopping),
yields a non-trivial sequence of quantum glass phases that 
intervene at the interface between the Mott insulating and the superfluid
regimes of the clean system. In particular, the coexistence of gapless and 
gapped regions close to the Mott insulating phase leads to a novel Mott glass 
phase which is incompressible and gapless. It shares 
some of the global properties of the Mott insulator, but resemble the Bose glass
in local properties. It is remarkable 
that the phase boundaries of the Mott insulating phase are basically 
unaffected by off-diagonal disorder, and that the extent of the Mott glass
can be tuned by varying the range of the spatial correlations 
in the disorder realization. It is also evident that particle-hole
symmetry is not a pre-requisite for observing the Mott glass phase. The
essential ingredients for its realization are off-diagonal (hopping)
disorder and commensurate filling. Finally we observe that while the
present results are obtained for a one-dimensional system, the Mott glass
phase is expected to be more robust in higher dimensions. 
As seen here,
the robustness of the Mott glass phase depends crucially on having large 
domains with uniform intra-domain hopping. This is achieved more
readily in higher dimensions due to increased co-ordination number.
Indeed, in three dimensions, the Mott glass phase is expected to persist to finite
temperatures. 

\noindent
\underbar{Acknowledgments} We thank T.~Roscilde for fruitful discussions. 
The work was supported by US DOE under Contract No. W-7405-ENG-36 (PS) and 
DE-FG02-05ER46240 (SH). S.H. ackowledges the hospitality of the Los Alamos 
National Laboratory where part of this work was carried out. The simulations 
were carried out at the high-performance computing computing center at USC.


\begin{thebibliography}{99}
\bibitem{footnote1}
In cold atom experiments, implementations of
off-diagonal disorder are accompanied by 
diagonal (potential) disorder. Although we consider only 
hopping-disorder explicitly, the
interaction and on-site (chemical) potential energies are expressed
in units of $t$ -- this results in
an implicit realization of diagonal disorder.
\bibitem{footnote2}
The finite width of the peaks is due to the
finite sizes of the domains, and thus increases with increasing
range of the spatial correlations in the disorder.
\end{thebibliography}
\end{document}